\documentstyle [12pt]{article}
\begin{document}
\centerline {Fourth International Workshop on Theoretical and 
Phenomenological Aspects}
\centerline {~~~~~~~~~~~of Underground Physics, Toledo (Spain) 
September 17-21 1995}
\vskip 1cm
\centerline {\bf COSMOLOGICAL IMPLICATIONS OF A POSSIBLE CLASS OF}
\vskip 3mm
\centerline {\bf PARTICLES ABLE TO TRAVEL FASTER THAN LIGHT}
\vskip 1.5cm
\centerline {\bf L. GONZALEZ-MESTRES}
\vskip 4mm
\centerline {Laboratoire d'Annecy-le-Vieux de Physique des 
Particules,}
\centerline {B.P. 110 , 74941 Annecy-le-Vieux Cedex, France}
\vskip 1mm
\centerline {and} 
\vskip 1mm
\centerline {Laboratoire de Physique Corpusculaire, 
Coll\`ege de France,}
\centerline {11 pl. Marcellin-Berthelot, 75231 Paris Cedex 05 , 
France}
\vskip 1.5cm
{\bf Abstract}
\vskip 2mm
We discuss the possible cosmological implications of a class of 
superluminal particles,
in a scenario where: a) Lorentz invariance is only an approximate 
property of the equations of a sector of matter;
b) several critical speeds of matter in vacuum exist.
The Big Bang scenario and the evolution of the very early universe, 
as well as large scale structure, can be strongly influenced by
the new particles.
\vskip 1cm
{\bf 1. THE NEW PARTICLES}
\vskip 5mm
In a recent paper [1], we proposed a new class of
non-tachyonic superluminal particles assuming the existence of several
critical speeds of matter in vacuum (like those of light and sound in a 
defectless, perfectly transparent crystal at zero temperature).
Considering an analogy with sine-Gordon solitons in a galilean world, 
we pointed out that the apparent Lorentz
invariance of the laws of physics does not imply that space-time is 
indeed mikowskian.
\vskip 4mm
Special relativity is usually presented as an intrinsic property of 
space-time and geometry is the
startpoint of the theory of gravitation in general relativity. 
However, a look to various dynamical systems
would suggest a more flexible view with the properties of matter 
playing the main role.
In a two-dimensional galilean space-time, a sine-Gordon soliton 
of the form: 
\begin{equation}
\phi _v(x,t)~=~4~arc~tan~[exp~(\pm~\omega c_0^{-1}z)]
\end{equation}
with:
\begin{equation}
z~=~(x~-~vt)~(1~-~v^2/c_0^2)^{-1/2}
\end{equation}
has speed $v$ ,
critical speed $c_0$ and all kinematical properties of a 
"relativistic particle", $c_0$ playing the role of the
speed of light.
An "absolute rest frame" would be provided by the rest frame of the 
dynamical system, but it cannot be felt by the soliton
unless the soliton is accelerated to a speed extremely close to $c_0$ 
(enough to feel the
small distance structure of the dynamical system).
Similarly, Lorentz invariance can be only an approximate property of 
equations describing a sector of matter above a given scale
and an absolute rest frame can exist without contradicting the 
minkowskian structure felt by ordinary particles. Then $c$ ,
the speed of light, will not necessarily be the only critical speed in 
vacuum. Superluminal sectors may exist
related to new degrees of freedom not yet unraveled experimentally.
\vskip 4mm
The new particles would not be tachyons: they may feel different 
minkowskian space-times with critical speeds $c_i \gg  c$
(the subscript $i$ stands for the $i$-th superluminal sector) and 
behave kinematically like ordinary particles apart from
the difference in critical speed. The "ordinary" sector will contain 
"ordinary" particles with critical speed $c$ .
Each sector will have its own Lorentz invariance, but interaction 
between two different sectors will break both
Lorentz invariances. 
Lorentz invariance will be
 simultaneoulsy apparent for several sectors in at best one 
inertial frame (the "vacuum rest frame", i.e. the "absolute" rest frame).
Superluminal particles will have [1] rest energy:
\begin{equation} 
E_{rest}~=~mc_i^2 
\end{equation}
for inertial mass $m$
and critical speed $c_i$ . Energy and momentum conservation will 
not be spoiled by the existence of several critical speeds in
vacuum: conservation laws will as usual hold for phenomena leaving 
the vacuum unchanged. If superluminal particles couple
weakly to ordinary matter, their effect on the ordinary sector may 
occur basically at
very high energy and short distance, far from
the domain of
conventional tests of Lorentz invariance. Thus, nuclear and particle 
physics experiments may open new windows in this field.
\vskip 4mm
At $v > c$ , the new superluminal particles
are kinematically allowed to emit "Cherenkov" radiation (on-shell 
ordinary particles) in vacuum; at $v > c_i$ , they can
emit particles of the $i$-th superluminal sector. "Cherenkov" 
radiation in vacuum provides a signature
for the production of superluminal particles at high energy 
accelerators, but the ability to radiate in vacuum will depend
on the produced particles.
Hadron colliders are the safest way to produce
superluminal particles, as quarks couple to all known interactions.
For the far future, one may think of a high energy
collider as the device to emit modulated and directional superluminal signals.
\vskip 4mm
Each superluminal sector may be protected by a conserved quantum number
and each sectorial "lightest superluminal particle" may be stable, 
but this is not unavoidable and we may be inside a sea of
long-lived superluminal particles which decay into ordinary 
particles or into lighter superluminal ones.
Such decays could play a cosmological role, and even be observable.
As the graviton is an "ordinary" gauge particle associated to the local Lorentz
invariance of the ordinary sector, superluminal particles will not 
couple to gravity in the usual way.
Each sector may generate its own "gravity" associated to the sectorial 
Lorentz invariance. Concepts so far
considered as very fundamental (i.e. the universality of the exact 
equivalence between inertial and gravitational mass)
will become approximate sectorial properties. 
\vskip 4mm
Even if such a scenario brings us somehow back to ether,
it is not in contradiction with modern particle physics where the 
vacuum is clearly not empty and has an important internal
structure which is just starting to be explored.
Gravitational properties of vacuum remain basically unknown and new
forces may have governed the expansion of the Universe.
\vskip 6mm
{\bf 2. COSMOLOGICAL IMPLICATIONS}
\vskip 5mm
If superluminal sectors exist and Lorentz invariance is only an approximate
sectorial property, the Big Bang scenario may become quite different, as:
a) Friedmann equations do no longer govern the global evolution of 
the Universe, which will be influenced
by new sectors of matter coupled to new forces and with different 
couplings to gravitation; b) gravitation itself
will be modified, and can even disappear, at distance scales where 
Lorentz invariance does no longer hold;
c) at these scales, extrapolation to a "Big Bang limit" from low 
energy scales does not make sense;
d) because of the degrees of freedom linked to superluminal sectors, 
the behaviour of vacuum will be different from standard
cosmology; e) the speed of light is no longer an upper limit to the 
speed of matter.
No basic consideration seems to prevent "ordinary" interactions 
other than gravitation from
coupling to the new dynamical sectors with their usual strengths. 
Conversely, ordinary particles can in principle couple to 
interactions mediated by superluminal objects.
Conventional covariant derivatives can
be used for all particles and gauge bosons independently of their 
critical speed in vacuum. However, experience seems to
suggest that superluminal particles have very large rest energies 
or couple very weakly to the ordinary sector.
\vskip 4mm
Each sectorial Lorentz invariance is expected to break down below a 
critical distance scale, when the Lorentz-invariant
equations (and Lorentz-covariant degrees of freedom) can no longer 
be used and a new dynamics appears.
For a sector with critical speed $c_i$ and apparent Lorentz 
invariance at distance
scales larger than $k_i^{-1}$ , where $k_i$ is a critical wave 
vector scale, we can expect
the appearance of a critical temperature $T_i$
given approximately by:
\begin{equation}
k~T_i~\approx ~\hbar ~c_i~k_i
\end{equation}
where $k$ is the Boltzmann constant and $\hbar $ the Planck constant. 
Above $T_i$ , the vacuum will not necessarily allow for the 
previously mentioned particles
of the $i$-th sector and new forms of matter can appear. 
If $k_0$ stands for the
critical wave vector scale of the ordinary sector,
above $T_0$ $\approx $ $k^{-1}\hbar ck_0$
the Universe may 
have contained only superluminal particles 
whereas superluminal and ordinary
particles coexist below $T_0$ . It may happen that some ordinary
particles exist above $T_0$ , but with different
properties (like sound above the melting point).
\vskip 4mm
If Lorentz invariance is not an absolute property of space-time, 
ordinary particles did most likely not govern the beginning of
the Big Bang (assuming such a limit exists)
and dynamical correlations have been able to propagate
must faster than light
in the very early Universe.
The existence of superluminal particles, and of the vacuum degrees 
of freedom which generate such excitations,
seems potentially able
to invalidate arguments leading to the
so-called "horizon problem" and "monopole problem", because:
a) above $T_0$ , particles and dynamical correlations are expected 
to propagate mainly at superluminal speed,
invalidating conventional estimates of the "horizon size";
b) below $T_0$ , the annihilation of superluminal
particles into ordinary ones is expected to release very large 
amounts of kinetic energy from the rest masses 
($E=mc_i^2$ , $c_i \gg c$) and generate
a fast expansion of the Universe. Conventional inflationary models 
rely on Friedmann equations and in principle cannot
hold in the new scenario. New inflationary models can be 
considered, but their need is far from obvious.
If a "Big Bang" limit exists,
and if
a generalized form of Friedmann equations can be written down 
incorporating all sectors and forces,
a definition of the horizon distance would be:
\begin{equation}
d_H(t)~=~R(t)~\int_0^t ~C(t') R(t')^{-1}~dt'
\end{equation}
where $d_H$ is the generalized horizon distance, $R(t)$ is the 
time-dependent cosmic scale factor and $C$ is the maximum of all
critical speeds. 
This definition is realistic even for ordinary particles, which can 
be radiated by superluminal ones at $v > c$ or produced by
their annihilation.
$C$ can be infinite if one of the sectors has no critical speed 
(as in the usual galilean space-time), or
if an infinite number of superluminal sectors exist.
The homogeneity and isotropy of our Universe, as manifested through 
COBE results, are now natural properties.
\vskip 4mm
The coupling between the ordinary sector and the superluminal ones
will influence black hole dynamics.
A detectable flux of magnetic monopoles (which can
even be superluminal) is not excluded, as the "horizon problem" can 
be eliminated without the standard inflationary scheme.
Long range correlations introduced by superluminal degrees of 
freedom can also play a role in the formation of
objects (i.e. strings) leading to large scale structure of the Universe.
\vskip 4mm
Physics at grand unified scales can present new interesting features.
Grand unified symmetries
are possible as sectorial symmetries of the vacuum degrees of 
freedom, even if ordinary particles do not exist
above $T_0$ . But analytic extrapolations (e.g. of running coupling 
constants) cannot be performed
above the phase transition temperature. 
If $kT_0$ is not higher
than $\approx $ 10$^{14}$ $GeV$
($k_0^{-1}$ $\approx $ 10$^{-27}$ $cm$ , time scale
$\approx $ 10$^{-38}$ $s$), the formation of a symmetry-breaking
condensate in vacuum 
may have occurred above $T_0$ and remain below the transition temperature.
Because of
superluminal degrees of freedom and of phase
transitions at $T_0$ and at all $T_i$ ,
it seems impossible to set a "natural time scale"
based on extrapolations from the low energy sector
(e.g. at the Planck time $t_p$ $\approx $ 10$^{-44}$ $s$ 
from Newton's constant).
Arguments leading to the "flatness" or "naturalness"
problem, as well as
the concept of the cosmological constant and the relation between 
critical density and Hubble's "constant" (one of the basic
arguments for dark matter at cosmic scale),
should
be reconsidered.
\vskip 4mm
At lower temperatures, superluminal particles do not disappear. 
In spite of limitations coming from
annihilation, decoupling and "Cherenkov radiation",
they can produce important effects in the evolution of the Universe.
Superluminal matter may presently be dark, with an unknown coupling 
to gravitation and coupled to
ordinary matter by new, unknown forces.
\vskip 6mm
{\bf Reference}
\vskip 2mm
[1] L. Gonzalez-Mestres, "Properties of a possible class of particles
able to travel faster than light", Proceedings of the
Moriond Workshop on "Dark Matter in Cosmology, Clocks and Tests of 
of Fundamental Laws", Villars (Switzerland), January 21-28 1995, 
Ed. Fronti\`eres. Paper astro-ph/9505117 .
\end{document}